\documentclass[fleqn,a4paper,12pt]{article}
\usepackage{amsmath,amssymb,amsfonts,amsthm}
\usepackage{color}
\usepackage{graphicx}
\usepackage{psfrag}
\usepackage{hyperref}
\usepackage{natbib}
\graphicspath{{./figs/}}

\usepackage{enumitem}

\usepackage{enumerate}
\numberwithin{equation}{section}

\newfont{\bbold}{msbm10 scaled\magstep1}

\RequirePackage{latexsym}

\renewcommand{\vec}[1]{\mbox{\boldmath $#1$}}

\newcommand{\FM}{\Phi} 

\begin{document}
\sffamily
\begin{flushleft}
\begin{center}%
\end{center}%
\vspace{.5cm}

{\Large {\bf Equilibria, periodic orbits and computing them.}}\\[4pt]

{\large EPSRC Summer School on Modal decompositions in fluid mechanics.}\\
{DAMTP, Cambridge, 5-8 August 2019.}

{\textbf{Ashley P.\ Willis}},\\
{\textbf{School of Mathematics and Statistics}}, 
{\textbf{University of Sheffield, U.K.}}\\%
a.p.willis@sheffield.ac.uk, \href{http://www.openpipeflow.org}{openpipeflow.org}.
\vspace{.5cm}
\hrule\vspace{.3cm}%
\end{flushleft}

In this short exposition, we describe equilibria and periodic
orbits in terms of the flow map, $\vec{\FM}$, and discuss the
essentials of the Jacobian-free Newton--Krylov (JFNK) method that can be 
used to find them.
This method requires little more than calls to an existing time stepping code, 
which $\vec{\FM}$ can be considered to represent. 
Fortran90\,/\,MATLAB code is available to try it out for yourself,
where, in the template/example the method is applied to the Lorenz system.
This code is problem-independent and can be applied to large systems, having
initially been developed to find periodic 
orbits in simulations of pipe flow.

\section{Using the flow-map}

Let the point $\vec{x}_0$ be an $n$-vector representing 
the state of a system.
For a dynamical system, as 
time $t$ progresses, the point $\vec{x}_t$ traces out a trajectory, 
a one-dimensional curve, 
in an $n$-dimensional phase space $\mathcal{M}$.  

We can describe the trajectory for $\vec{x}_t$ using the 
{\bf flow-map}
denoted $\vec{\FM}^t$, which
takes a point $\vec{x}_0$ and evolves it by a time period $t$:
~~
$  \vec{\FM}^t : \vec{x}_0 \to \vec{x}_t $ ~~i.e.
\begin{equation} \label{eq:flow}
  \vec{x}_t = \vec{\FM}^t(\vec{x}_0)\, .
\end{equation}
More generally, the flow-map simply advances a point along its trajectory:
\begin{equation}
   \vec{x}_{t'+t} = \vec{\FM}^t(\vec{x}_{t'}) \, .
   \qquad \mbox{\textcolor{blue}{[flow-map]}}
\end{equation}

\subsection{Example: Lorenz's model for convection}

\begin{figure}[t!]
\centering
(a)
\includegraphics[width=70mm]{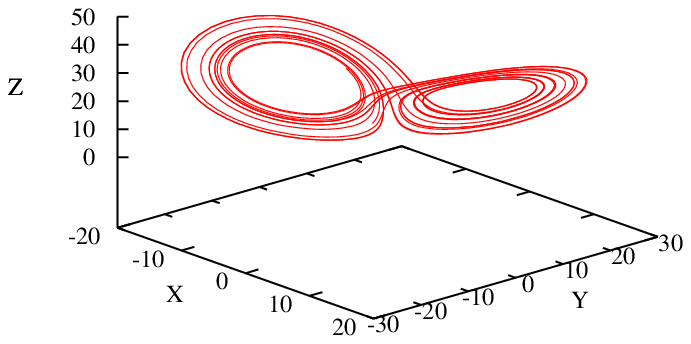}
~~~
\psfrag{FM18}{\small $\vec{x}_{18}$}
\psfrag{FM19}{\small $\vec{x}_{19.6}=\vec{\FM}^{1.6}(\vec{x}_{18})$}
(b)\includegraphics[width=70mm]{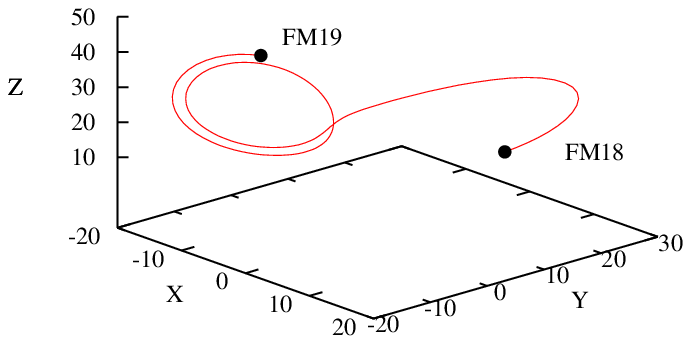}
\caption{ \label{fig:lorenzflowmap} 
(a) Lorenz attractor.  (b) The flow-map $\vec{\FM}$ is used to
advance the state $\vec{x}$ by 1.6 time units from $t=18$ to $t=19.6$.
}
\end{figure}

\cite{lorenz63} derived the following system for three-time dependent
amplitudes, $X(t)$, $Y(t)$ and $Z(t)$:
\begin{eqnarray} \label{eq:Lorenz63}
   \dot{X} &=& ~~~~~~~~-\sigma\, X + \sigma\, Y \, , \nonumber \\
   \dot{Y} &=& -X\,Z + r\,X -~~ Y \, ,\\
   \dot{Z} &=& ~~X\,Y ~~~~~~~~~~~~~~~~~- b\,Z \, . \nonumber 
\end{eqnarray}
At each instant in time, the current state 
$\vec{x}=(X,Y,Z)$ is a point in the phase space 
$\mathcal{M}=\mathbb{R}^3$.  
As time progresses, $\vec{x}_t=(X(t),Y(t),Z(t))$ traces out a trajectory,
i.e.\ a curve, in $\mathbb{R}^3$.
Lorenz focussed on 
parameter values 
$r=28$, $b=8/3$, $\sigma=10$, which result in chaotic trajectories.
The flow-map takes us along this trajectory, see figure 
\ref{fig:lorenzflowmap}.

\begin{figure}[t!]
\centering
\psfrag{Cold}{\small Cold}
\psfrag{Hot}{\small Hot}
\psfrag{H}{$H$}
\psfrag{x}{$x$}
\psfrag{z}{$z$}
\psfrag{L}{$\lambda$}
\vspace{1mm}
\includegraphics[width=60mm]{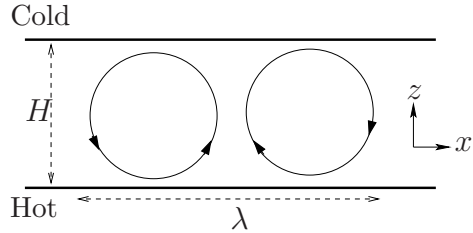}
\vspace{-3mm}
\caption{ \label{fig:RBconv} 
Rayleigh--B\'enard convection.  
A pair of convection rolls,
wavelength $\lambda=(2/a)\,H$; $b=4/(1+a^2)$.
}
\end{figure}

We should not forget that each point $\vec{x}_t$ in phase-space 
corresponds to a whole convection flow pattern!  
Here it corresponds to a two-dimensional flow between
two flat plates a distance $H$ apart, 
with a temperature difference $\Delta T$
between the top and bottom plates (figure \ref{fig:RBconv}).
The amplitudes $X(t)$, $Y(t)$, $Z(t)$ correspond to modulated 
variations in the temperature and velocity fields:
\[
   T(x,z,t) = \theta(x,z,t) - (z/H)\,\Delta T,
\qquad
u_x = - \frac{\partial \psi}{\partial z}, \quad
      u_z =  \frac{\partial \psi}{\partial x} \, , ~~~
      \mbox{(stream function $\psi$)}
\]
\vspace{-10mm}
\begin{eqnarray}
      \psi &=& X(t) \times \sin(\pi a x/H)\sin(\pi z/H) 
\times c_1,
      \nonumber \\[2pt]
      \theta &=& Y(t) \times \cos(\pi a x/H)\sin(\pi z/H) \times c_2
    ~-~ Z(t) \times \sin(2\pi z/H) \times c_3, 
      \nonumber
\end{eqnarray}
where the $c_i$ are scalar constants.


\vspace{-5mm}
\subsection{Invariant solutions}
\label{sect:invsolns}

Equilibria and periodic orbits are topological features of the 
phase space $\mathcal{M}$.  Irrespective of the coordinates
and measure of distance used to visualise the phase space, an 
equilibrium remains a point, and a periodic orbit remains a closed
loop.  Properties, such as their eigenvalues, are also {\em invariant}
to the measure of distance used.

An {\bf equilibrium} $\vec{x}_0$ is a {\bf fixed point} of the 
flow-map that satisfies
\begin{equation}
  \label{eq:eqm}
  \vec{x}_0 = \vec{\FM}^t(\vec{x}_0)\, ,
  \qquad \mbox{\textcolor{blue}{[equilibrium = fixed point]}}
\end{equation}
for {\em any} time $t$.  
A point $\vec{x}_p$ on a {\bf periodic orbit} satisfies
\begin{equation} \label{eq:PO}
  \vec{x}_p = \vec{\FM}^T(\vec{x}_p)\, ,
  \qquad \mbox{\textcolor{blue}{[periodic orbit]}}
\end{equation}
where $T$ is the period of the orbit.
In terms of solutions of the flow-map, we can consider an equilibrium 
to be a special case of a periodic orbit where $T$ may be arbitrarily
chosen.

\begin{figure}[h!]
\centering
(a)
\psfrag{0}{\small $0$}
\psfrag{l}{\small $l=ct$}
\psfrag{x0}{\small $\vec{x}_0$}
\psfrag{xt}{\small $\vec{x}_t$}
\includegraphics[width=45mm]{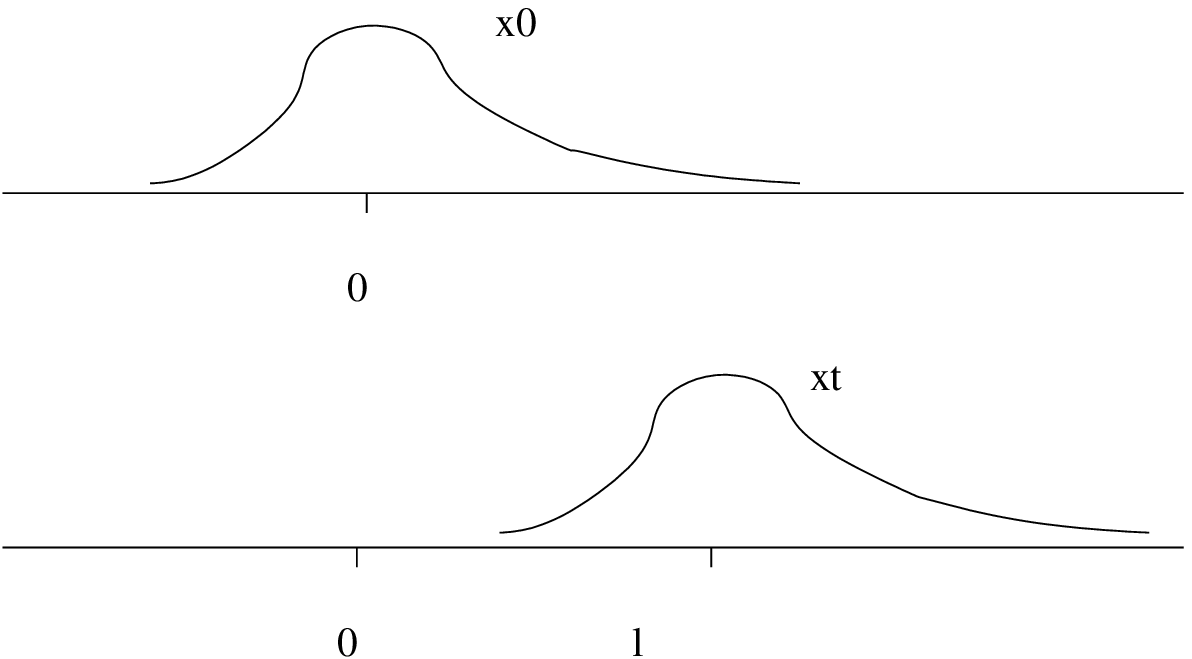}
~~~~~~
(b)
\psfrag{0}{\small $0$}
\psfrag{x0}{\small $\vec{x}_0$}
\psfrag{xt}{\small $\vec{x}_{T/2}$}
\includegraphics[width=45mm]{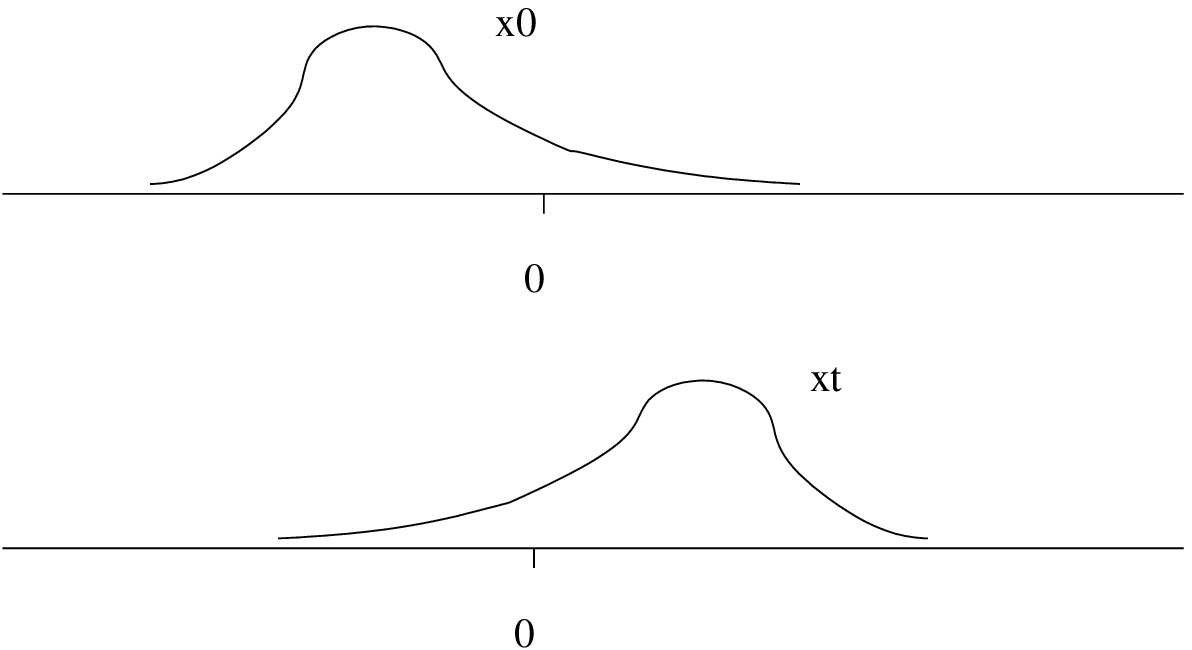}
\vspace{-2mm}
\caption{ \label{fig:TWs} 
(a)
A travelling wave: $\vec{x}_t=g(ct)\,\vec{x}_0$.  By shifting back, $\vec{x}_0=g(-l)\,\vec{x}_t=g(-l)\,\vec{\FM}^t(\vec{x})$.
(b) A pre-periodic orbit: $\vec{x}_0=\vec{x}_T$, but also
$\vec{x}_0=\sigma\,\vec{x}_{T/2}$, 
where $\sigma$ flips the state about $0$.
}
\end{figure}

If a system has a homogeneous dimension, $x$, then it can have
{\bf travelling wave} solutions. (See figure \ref{fig:TWs}a.)
In a frame moving at some phase speed $c$, the solution looks
steady.  Equivalently, we can keep shifting the solution so
that it looks steady.  Let $g(l)$ be an operator that
shifts a state by a distance $l$ in the $x$ direction.  Then,
a travelling wave satisfies
\begin{equation}
  \vec{x}_0 = g(-ct)\,\vec{\FM}^t(\vec{x}_0)\, ,
  \qquad \mbox{\textcolor{blue}{[relative equilibrium = travelling wave]}}
\end{equation}
for any time $t$.  The state $\vec{x}_0$ is an equilibrium
solution of the slightly modified map for any $t$, hence 
we also call a travelling wave a {\bf relative equilibrium}.
Similarly, a periodic solution that recurs up to a spatial shift,
\begin{equation}
  \vec{x}_p = g(-\bar{c}T)\,\vec{\FM}^T(\vec{x}_p)\, ,
  \qquad \mbox{\textcolor{blue}{[relative periodic orbit]}}
\end{equation}
for some $\bar{c}$, we call a {\bf relative periodic orbit}.

\begin{figure}[h!]
\centering
\vspace{-5mm}
\includegraphics[width=44mm]{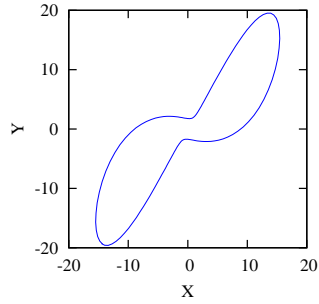}
\vspace{-5mm}
\caption{\label{fig:LozABrot}
 The shortest periodic orbit of the Lorenz system (\ref{eq:Lorenz63})
has the rotational symmetry 
$(X,Y,Z)\to(-X,-Y,Z)$.
}
\end{figure}

Suppose the dimension $x$ has a mirror symmetry about $x=0$.
Let $\sigma$ be the flip operator:
 $\sigma\,\vec{x}(x)=\vec{x}(-x)$.
The second half of an orbit satisfying (\ref{eq:PO}) might just
be a reflection of the first half:
\begin{equation} \label{eq:flip}
 \vec{x}_p = \sigma\,\vec{\FM}^{T/2}(\vec{x}_p)\,.
  \qquad \mbox{\textcolor{blue}{[pre-periodic orbit]}}
\end{equation}
(See figure \ref{fig:TWs}b.)
Such orbits are called {\bf pre-periodic orbits.}  The shortest/simplest
periodic orbits of a system with discrete symmetries 
are typically pre-periodic.  
Figure \ref{fig:LozABrot} shows  
the shortest periodic orbit of the Lorenz system,
where the second half of the orbit is related to the 
first half by the 180 degree rotation symmetry $(X,Y,Z)\to(-X,-Y,Z)$.


\newpage
\subsection{Poincar\'e sections}
\label{sect:Poincare}

\begin{figure}[h!]
\centering
\psfrag{P}{\small $\mathcal{P}$}
\psfrag{x'}{\small $\vec{x}'$}
\psfrag{t'}{\small $\vec{t}'$}
\psfrag{xp}{\small $\vec{x}_p$}
(a)\includegraphics[height=36mm]{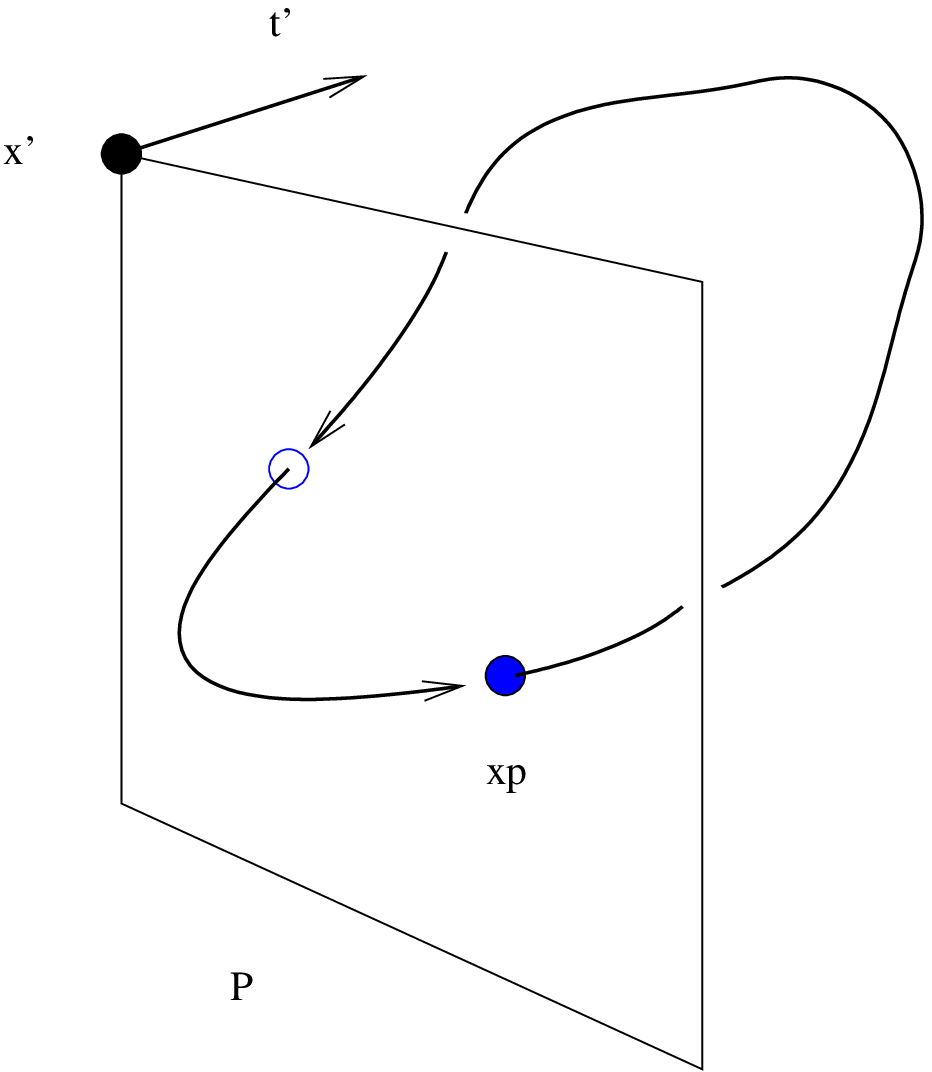}
~~~~~~
\psfrag{Mhat}{\small $\hat{\mathcal{M}}$}
\psfrag{x'}{\small $\vec{x}'$}
\psfrag{xhat}{\small $\hat{\vec{x}}_t$}
\psfrag{xt}{\small $\vec{x}_t$}
\psfrag{xt+T}{\small $\vec{x}_{t+T}$}
(b)\includegraphics[height=36mm]{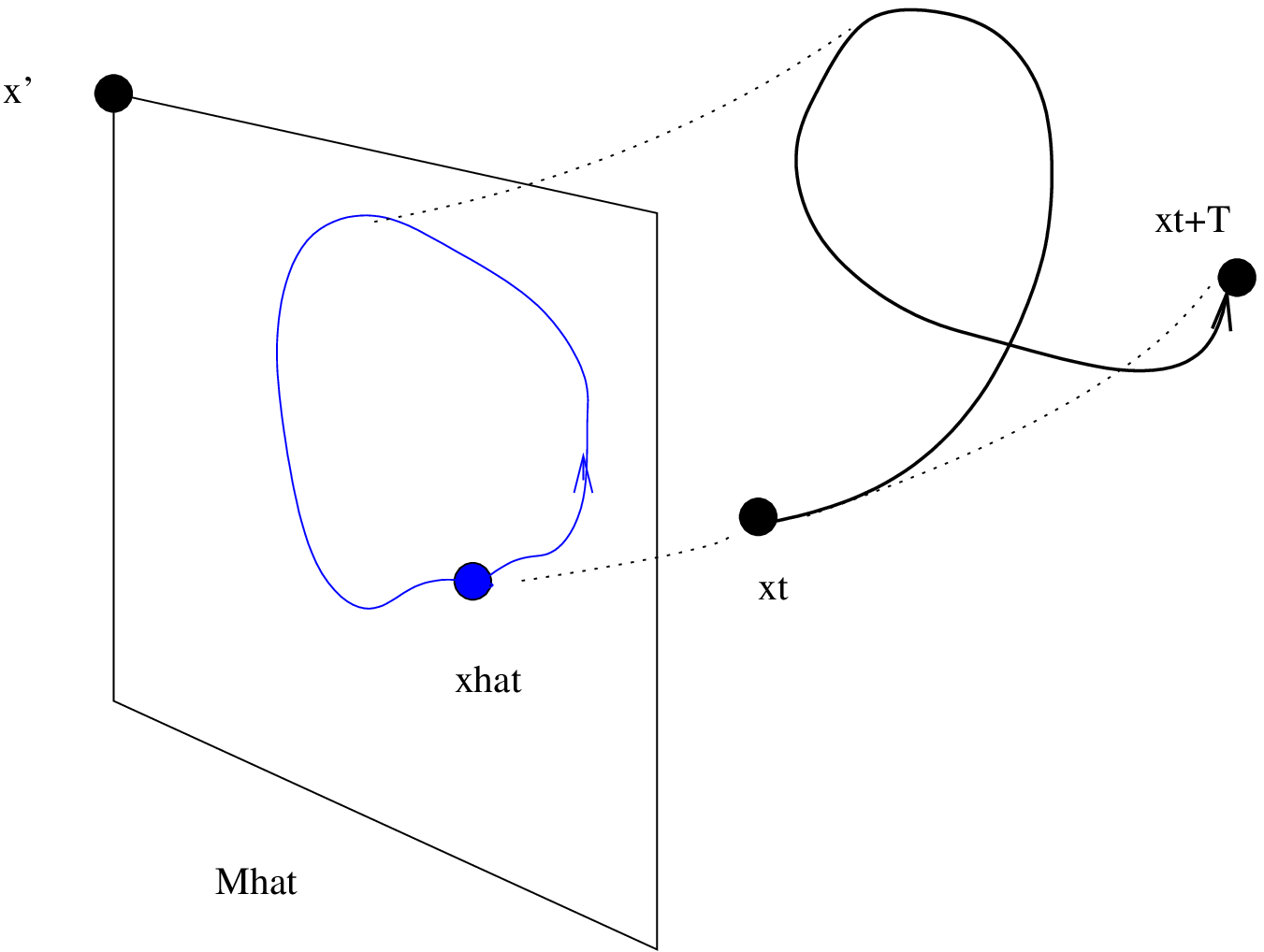}
\vspace{-1mm}
\caption{ \label{fig:Poincaresect} 
(a)
A \textcolor{blue}{Poincar\'e section $\mathcal{P}$}, 
defined by a point $\vec{x}'$ and a normal
vector $\vec{t}'$, is pierced by a periodic orbit at the periodic
point $\vec{x}_p$.
(b)
The projection of a relative periodic orbit onto a 
\textcolor{blue}{slice $\hat{\mathcal{M}}$},
is a periodic orbit, $\hat{\vec{x}}_t=\hat{\vec{x}}_{t+T}.$
The whole orbit is projected onto $\hat{\mathcal{M}}$.
Each state $\vec{x}_t$ is projected by 
applying shifts along the dotted lines onto $\hat{\mathcal{M}}$.
}
\end{figure}

\vspace{-2mm}
Let $\vec{x}'$ be a point and $\vec{t}'$ be a normal vector
that together define a hypersurface $\mathcal{P}$.  Crossings 
of $\mathcal{P}$ can be defined by times $t$ when 
$\langle \vec{x}_t-\vec{x}'|\vec{t}'\rangle = 0$.
(See figure \ref{fig:Poincaresect}a.)
We might restrict to when crossings occur in a particular direction,
say when the inner product goes from negative to positive.

We can now let $\vec{\FM}$ be the map that takes one point on 
$\mathcal{P}$ to the next crossing point on $\mathcal{P}$.  
If a periodic orbit has a point $\vec{x}_p$ on $\mathcal{P}$,
then it satisfies
\begin{equation}
  \vec{x}_p = \vec{\FM}(\vec{x}_p)\, .
\end{equation}
%
The advantage is that we no longer need to worry about the period $T$ 
for periodic orbits.  The disadvantage is that we know 
nothing about what happens to the orbit off $\mathcal{P}$, and 
in general, not all periodic orbits cross a single $\mathcal{P}$.

\vspace{-3mm}
\subsection{Slicing}
\label{sect:slicing}

For a homogeneous spatial dimension $x$, the freedom of a 
pattern to appear at any location
is awkward when we want to compare states.
Slicing is an automatic shifting procedure
that removes this degree of freedom.

Here we will discuss the simplest form of slicing -- `Fourier' slicing.
\citep[See][]{BudCvi14,WiShCv15}
When a system has a homogeneous dimension, it is commonplace
to work with a periodic domain of length $L=2\pi/\alpha$.
In this case, construct
$
\vec{x}'=\vec{x}_c\cos\alpha x + \vec{x}_s\sin\alpha x,
$
where $\vec{x}_c$ and $\vec{x}_s$ are arbitrary states independent of 
$x$; they must not both be zero.

\begin{figure}
\centering
\vspace{2mm}
(a)
\hspace{-2mm}
\psfrag{a1a2}{\small $\vec{x}~\to~(a_1,a_2)$}
\psfrag{xh}{ {\small $\hat{\vec{x}}$}}
\psfrag{th}{\small $\theta$}
\psfrag{gx}{\small $g(l)\,\vec{x}$, for all $l$}
\includegraphics[height=28mm]{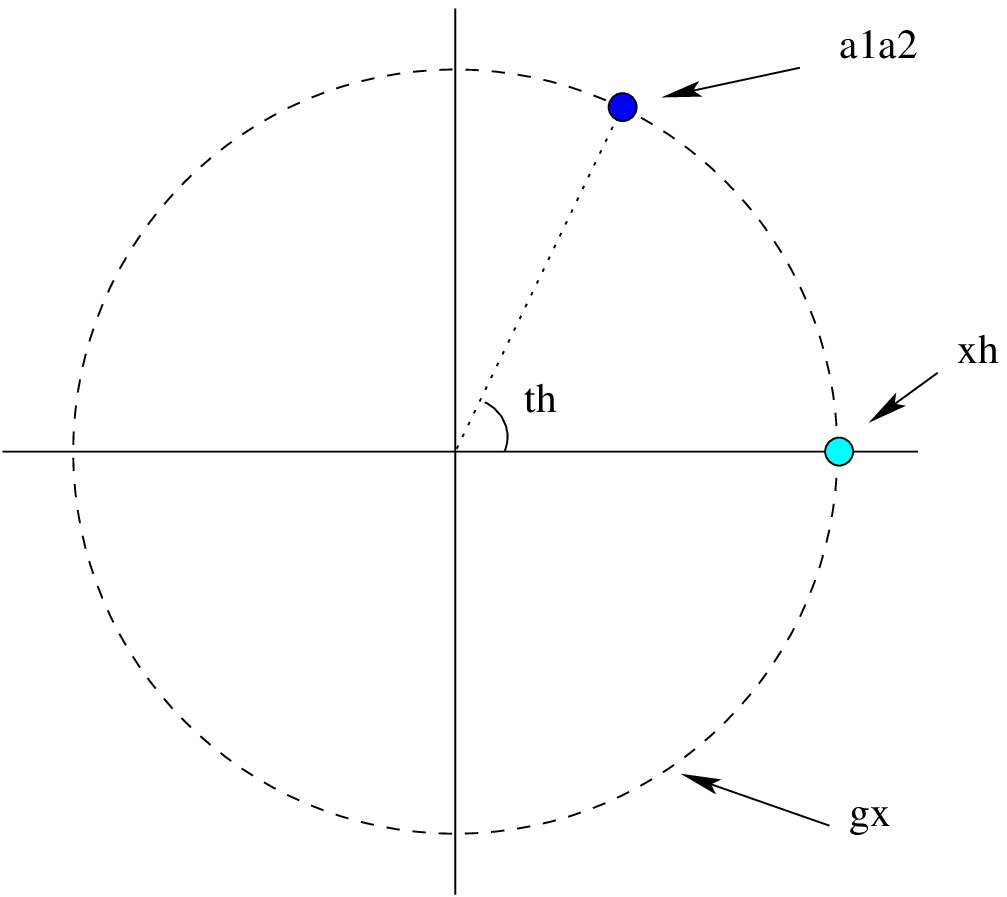}
\hspace{25mm}(b)
\psfrag{0}{\small $0$}
\psfrag{l0}{\small $l_0$}
\psfrag{x0}{\small $\vec{x}_0$}
\psfrag{xt}{\small $\vec{x}_t$}
\psfrag{lt}{\small $l_t$}
\psfrag{xh0}{\small $\hat{\vec{x}}_0$}
\psfrag{xht}{\small $\hat{\vec{x}}_t$}
\includegraphics[width=68mm]{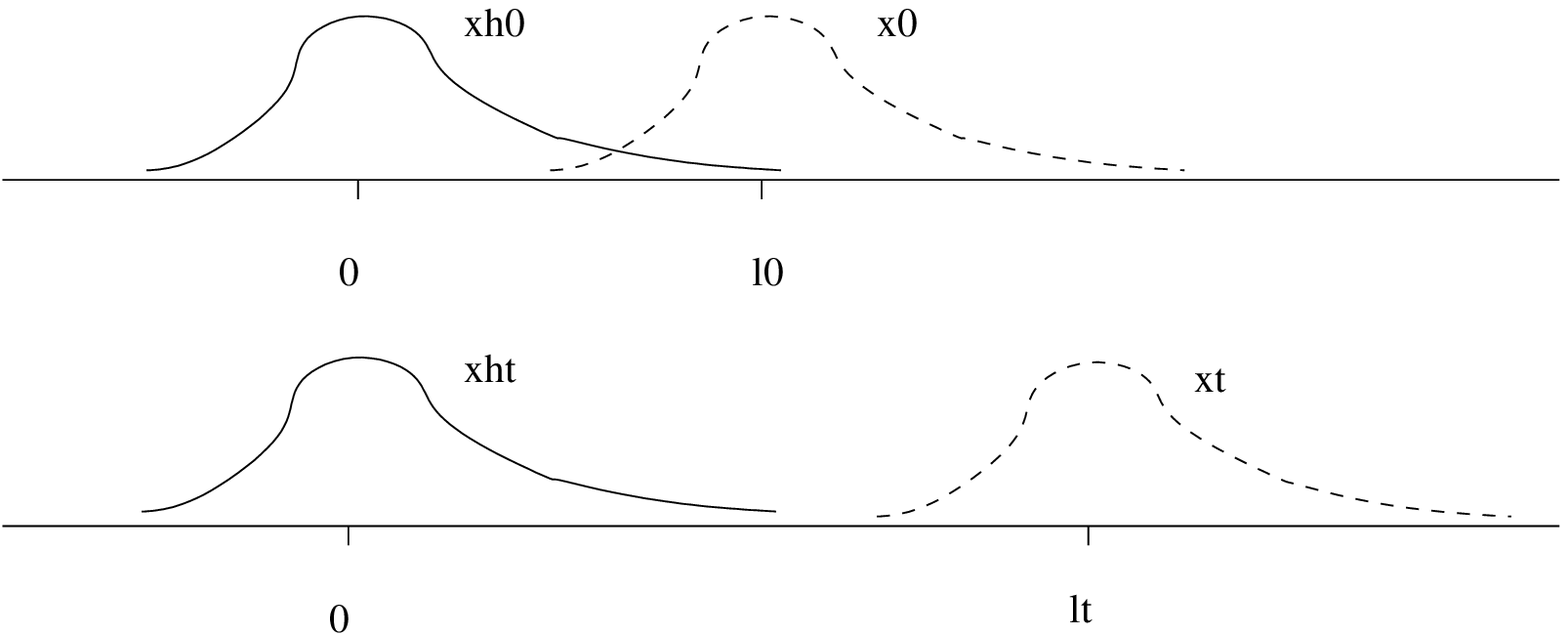}
\caption{ \label{fig:Fslice} 
(a)
Fourier slicing:  A state $\vec{x}$ is mapped onto the 
$(a_1,a_2)$-plane, where all shifted versions of $\vec{x}$ map onto a circle.  All versions are 
reduced 
to the single copy 
$\hat{\vec{x}}=g(-(\theta/2\pi)L)\,\vec{x}$.  
(b) A travelling
wave is reduced to an equilibrium.  $l_t$ is automatically determined by the slicing algorithm, from which a phase speed $c$ can be inferred.}
\end{figure}

Any state $\vec{x}$ may be projected onto a plane $(a_1,a_2)$ 
via $a_1=\langle\vec{x}|\vec{x}'\rangle$ and
$a_2=\langle\vec{x}\,|\,g(L/4)\,\vec{x}'\rangle$.
(See figure \ref{fig:Fslice}a.)
In this projection, the set of shifted states $\{g(l)\vec{x}\mbox{ for all }l\}$ lie on a circle centred on the origin.  
By shifting the states, we can rotate all points on the circle
to the {\em unique} point on the circle where it crosses the positive 
$a_1$ axis.  All possible shifted versions of $\vec{x}$ are then
mapped to the unique version
$\hat{\vec{x}}=g(-l)\,\vec{x}$, where $l=(\theta/2\pi)L$ and 
$\theta$ is the polar angle to $(a_1,a_2)$.

This operation is a {\bf symmetry-reduction}, 
and we say that the symmetry-reduced state $\hat{\vec{x}}$,
indicated by the hat, lies on a {\bf slice}.  
Arbitrary shifts have been eliminated, so the slice has dimension one 
less than that of the original system.  The slice $\hat{\mathcal{M}}$
is a hypersurface within the original space of states $\mathcal{M}$.

The slice is different from a Poincar\'e section because the symmetry
reduction can be applied to $\vec{x}_t$ for all times $t$.  We can
compute sliced dynamics with trajectories $\hat{\vec{x}}_t$
that lie within the slice.
(See figure \ref{fig:Poincaresect}b.)
Meanwhile, 
trajectories only pierce a Poincar\'e section.

{\bf Relative equilibria
(travelling waves) are reduced to equilibria} automatically:
\begin{equation}
   \vec{x}_0 = g(-ct)\,\vec{\FM}^t(\vec{x}_0)
   ~~   \to ~~
   \hat{\vec{x}}_0 = \hat{\vec{\FM}}^t(\hat{\vec{x}}_0) \, .
\end{equation}
All possible shifts of a state are reduced by shifting 
to one particular version on the slice $\hat{\mathcal{M}}$,
i.e.\ the travelling wave is `pinned' by the shifting.
(See figure \ref{fig:Fslice}b.)
Here, $\hat{\vec{\FM}}$ is the flow-map of the symmetry reduced
dynamics.  Note that {\em all} travelling waves of the system
are reduced to equilibria, even though they typically will
have a range of different phase speeds $c$.

Similarly, 
{\bf a relative periodic orbit becomes a closed periodic orbit},
because the start and end point are shifted to a single point on
$\hat{\mathcal{M}}$.  The relative periodic orbit
now satisfies the simpler form:
\begin{equation}
  \vec{x}_p = g(-\bar{c}T)\,\vec{\FM}^T(\vec{x}_p)
   ~~   \to ~~
  \hat{\vec{x}}_p = \hat{\vec{\FM}}^T(\hat{\vec{x}}_p)\, ,
\end{equation}
for any symmetry-reduced point $\hat{\vec{x}}_p$ on
the orbit.
%

\newpage
\section{Periodic orbits}

All periodic orbits (POs) in a chaotic attractor
must be unstable, otherwise the behaviour would eventually
be attracted to the orbit and become periodic, not chaotic.
So why is it useful to find POs when they're all unstable!? 

\vspace{-3mm}
\subsection{Why periodic orbits?}

\begin{figure}[h!]
 \centering
 \psfrag{x0}{$\vec{x}_0$}
 \psfrag{xp1}{$\vec{x}_{p'}$}
 \psfrag{xp2}{$\vec{x}_{p''}$}
 \includegraphics[width=60mm]{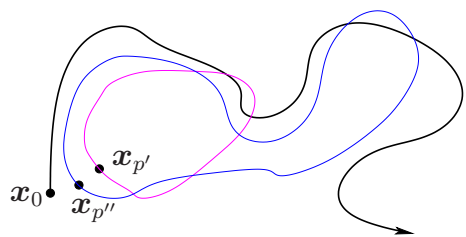}
 \vspace{-2mm}
 \caption{ \label{fig:shadowing}
   `Shadowing' of a chaotic trajectory by periodic orbits.  
   By considering periodic orbits of increasing length, it is possible
   to find a periodic point $\vec{x}_p$ arbitrarily close to a point
   $\vec{x}_0$ on the chaotic attractor.
}
\end{figure}

Firstly, the dynamics is always very close to a PO!
A chaotic attractor is {\bf dense} in POs: 
For any chaotic point $\vec{x}_0$ and $0<\epsilon\ll1$, 
there exists a periodic point $\vec{x}_p$ within $\epsilon$ of $\vec{x}_0$.
%
%
See $\mbox{figure \ref{fig:shadowing}}$.  Also...
%
\setlist{nolistsep}
\begin{itemize}[noitemsep]
\item
Unlike equilibria, POs exhibit dynamics! --- they capture the 
time-dependent dynamic processes of the system and organise
the chaotic set.
\item
The chaotic dynamics tends to follow the least unstable POs.
\item
A PO is a closed loop in
state space.  It will appear as a closed loop irrespective of the 
coordinates used to visualise the state space.  
\item
The shortest, most fundamental, POs provide an 
alphabet for symbolic dynamics.
\item
Statistical properties of a chaotic attractor can be calculated in 
terms of sums over the POs, using their relative stability.  
\end{itemize}

\vspace{-3mm}
\subsection{Examples of periodic orbits}

\begin{itemize}[leftmargin=*]
\item
The logistic map $x_{t+1} = r\,x_t\,(1-x_t)$ is chaotic 
for the case $r=4$ but has the (unstable) period-$2$ orbit
\[
(5-\sqrt{5})/8 ~\to~ (5+\sqrt{5})/8 ~\to~ (5-\sqrt{5})/8 ~\to~ ...
\]
\item
In a remarkable methodical and computational feat, 
\cite{DV03}
calculated 
111011 periodic orbits of the Lorenz attractor, that is
all periodic orbits with itineraries of up to length 20
(number of windings around the left/right `wings'),
with an accuracy of 14 decimal digits.
A few of them are shown in figure \ref{fig:vis03}.
The periodic orbit $A^{14}B$ was found to 
be the least unstable among all orbits calculated.
(It is the periodic orbit with the smallest Lyapunov/Floquet exponent.)
\begin{figure}[h]
 \centering
 \includegraphics[width=60mm]{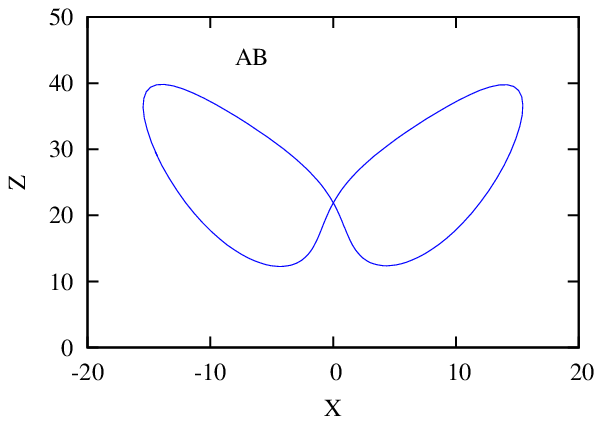}
 \includegraphics[width=60mm]{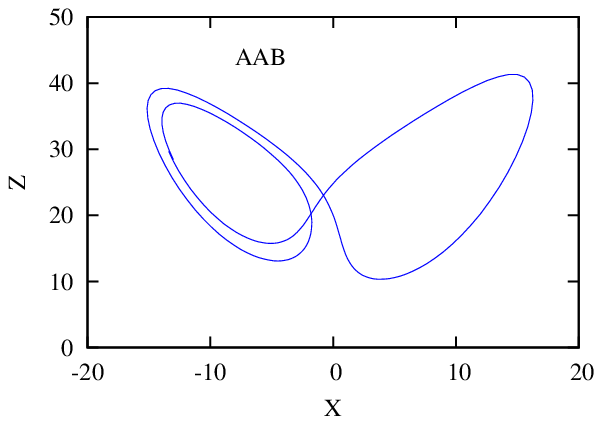}\\
 \includegraphics[width=60mm]{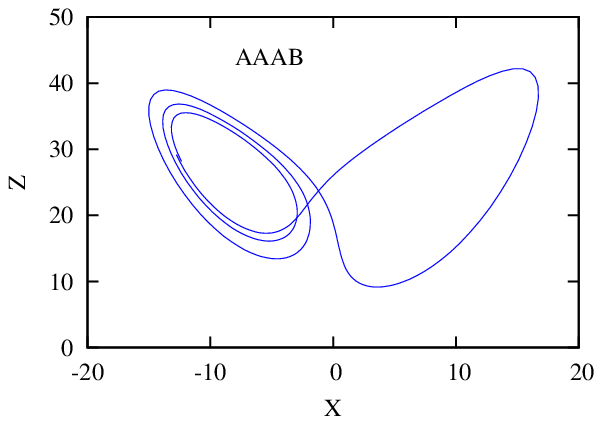}
 \includegraphics[width=60mm]{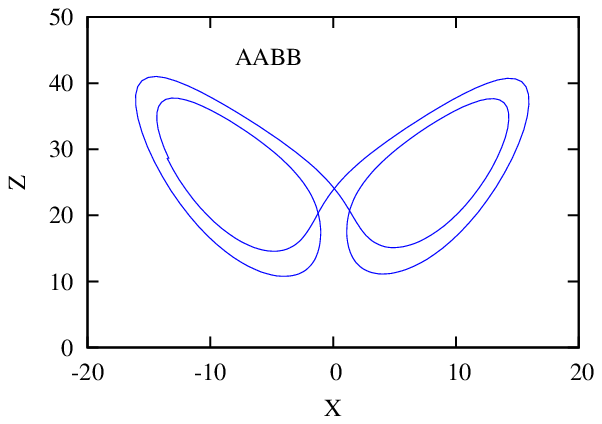}
 \caption{ \label{fig:vis03}
   A few of the shortest periodic orbits of the Lorenz system.
   \citep[Reproduction of figure 3 of][]{DV03}
}
\end{figure}
\item
For an $n=$\,154755-dimensional model of turbulent flow 
in a pipe,
\cite{WiShCv15}
calculated 
periodic orbits and
travelling wave solutions.
Travelling waves are stationary in a moving frame, 
so they correspond to fixed points in the sliced phase space.
Two `clouds' of solutions were observed, shown in figure \ref{fig:WSC16},
and the relationship between them found to be the reflection symmetry.
The phase-space visualisation reveals that trajectories don't
like to switch orientation with respect to the symmetry,
due to the presence of a strongly repelling (unstable)
fixed point, marked `A', that lives between the two clouds.
For beautiful examples from Couette flow, see \cite{CviGib10}.
\begin{figure}[h]
 \centering
 \includegraphics[width=60mm]{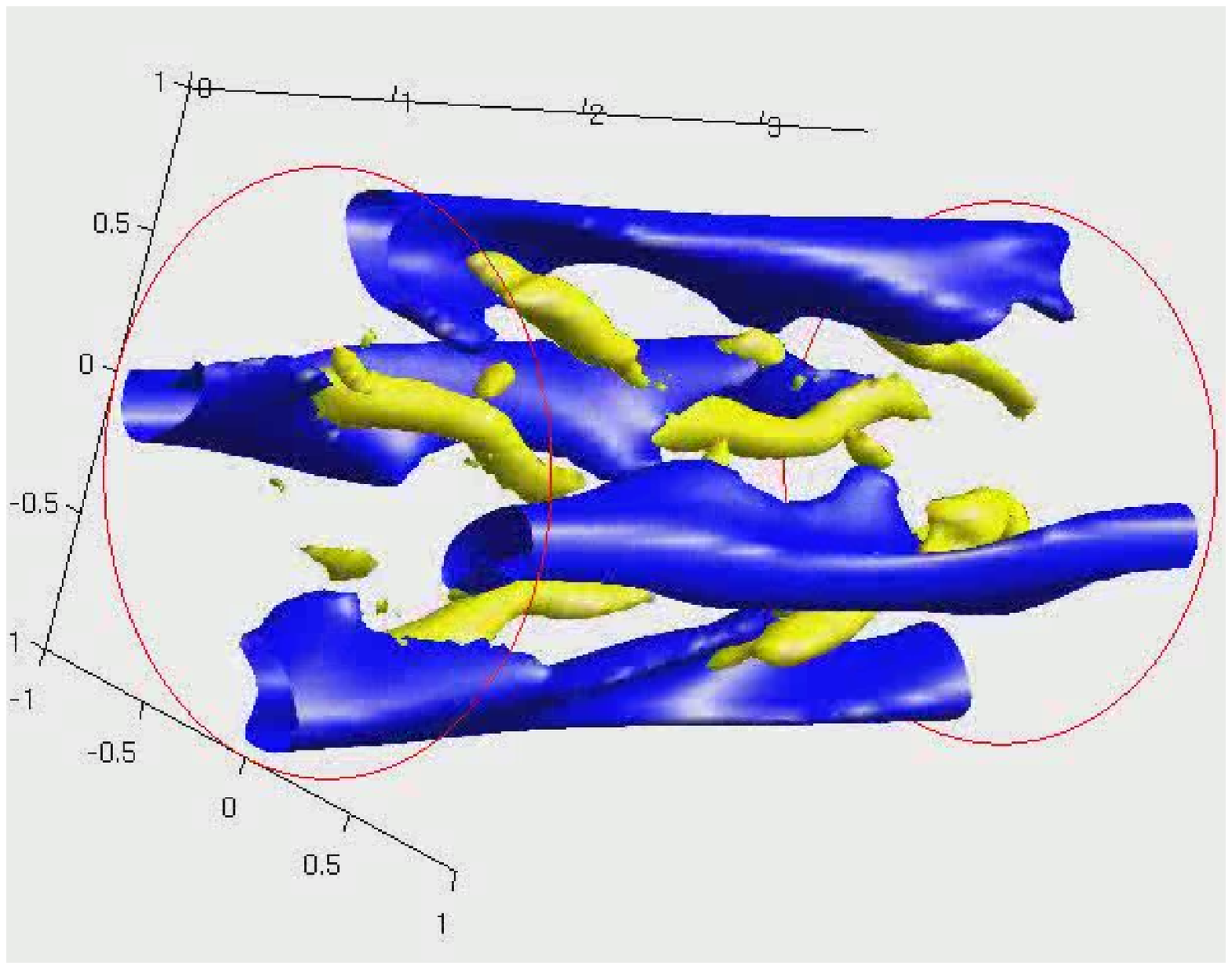} 
 \hspace{4mm}
 \includegraphics[width=75mm]{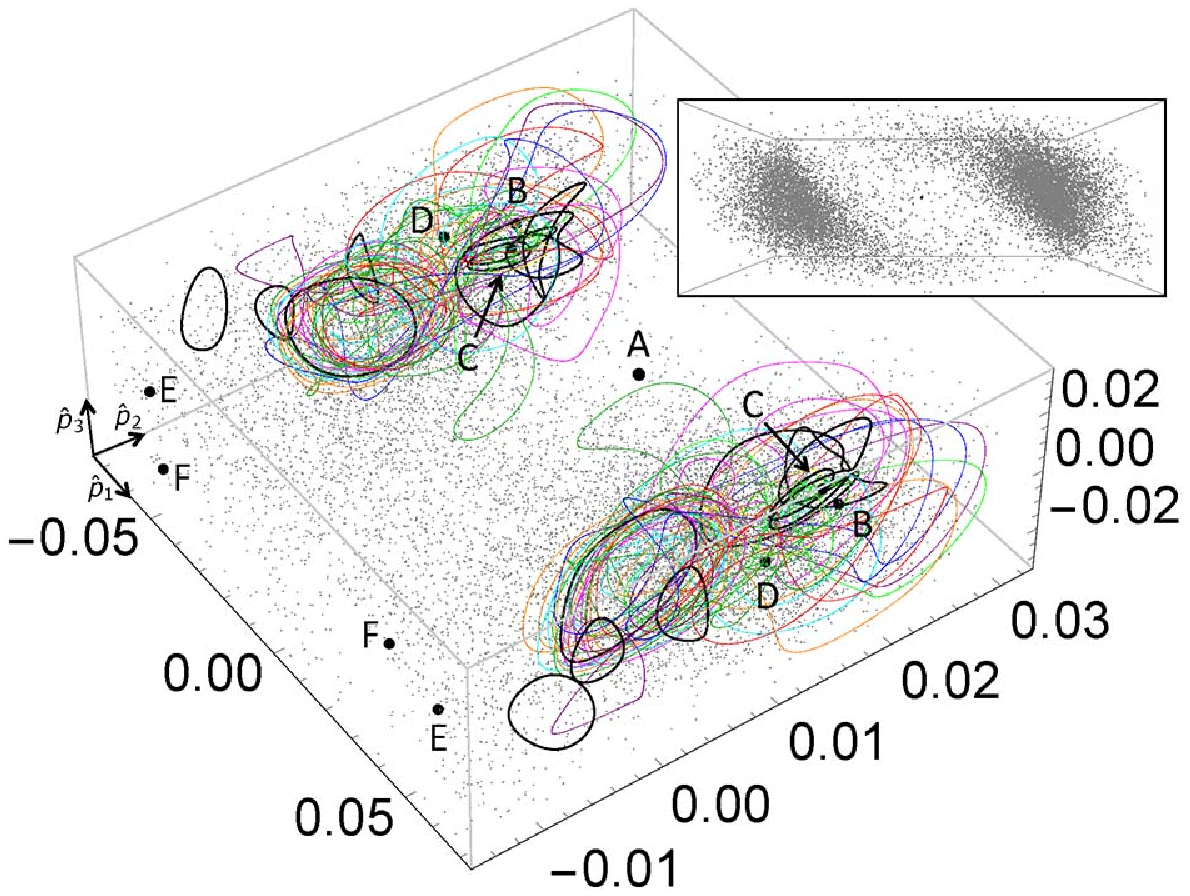} 
 \caption{ \label{fig:WSC16}
   ({\it left})
   Visualisation of pipe flow with slow streaks (blue) and vortices (yellow).
   ({\it right})
   Periodic orbits of pipe flow.  Inset is view from bottom left side of box.
   \citep[Reproduction of figure 4 of][]{WiShCv15}
}
\end{figure}
\end{itemize}

\subsection{Searching for recurrences}

At present, the standard approach to finding near-recurrences
is pretty crude.  It involves recurrence plots, with distance measures
tailored for the case at hand, and, if not automated,
visual inspection of the plot for close recurrences.
Nevertheless, this is the usual means to 
find points close to periodic orbits that can be refined
to exact recurrences using the Newton method (described in 
the next section).

\begin{figure}[b!]
 \centering
 \includegraphics[width=100mm]{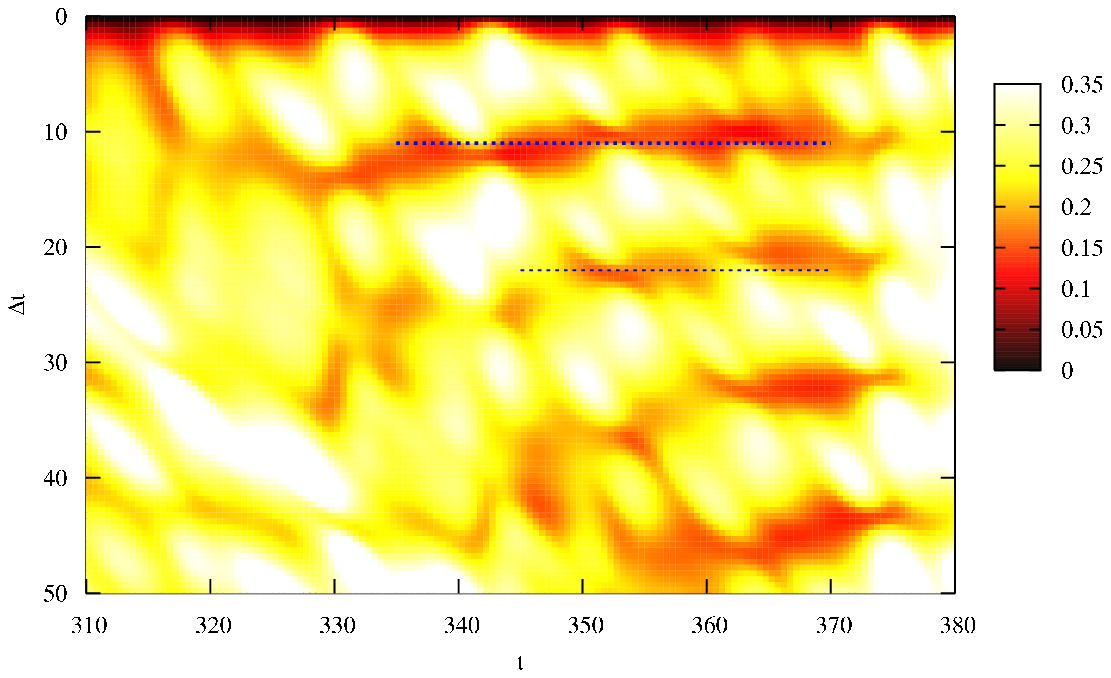}
\caption{ \label{fig:recurrence}
 Search for recurrences: colour plot of 
 $|| \hat{\vec{x}}_t - \hat{\vec{x}}_{t-\Delta t}||_c\,/\,||\hat{\vec{x}}_{t-\Delta t}||_c\, $ for pipe flow in a `minimal' box
\citep{ACHKW11}.
Axial shifts have 
been eliminated by slicing, indicated by the hat; section \ref{sect:slicing}.
Minima around the horizontal lines suggest shadowing of a periodic orbit
with period $T\approx11$.  Subscript $c$ indicates that compensation has 
been applied to pick up a signal from cross-flow components, which are 
smaller in magnitude but as important as the streamwise perturbations.
}
\end{figure}

For a recurrence plot we compute something like
\begin{equation}
   \frac{|| \vec{x}_t - \vec{x}_{t-\Delta t} || 
     }{||\vec{x}_{t-\Delta t}|| } \, .
\end{equation}
Figure \ref{fig:recurrence} is an example for pipe flow.
We then look for local minima in the plot that provide candidate
recurrent points, $\vec{x}_p\approx\vec{x}_{t-\Delta t}$ and $T\approx\Delta t$.
The normalisation factor might be chosen
to depend on both $||\vec{x}_t||$ and $||\vec{x}_{t-\Delta t}||$,
or might not be necessary at all.
A complication is that we might need to minimise over discrete 
symmetries, such as the flip operator $\sigma$, or over shifts
(section \ref{sect:invsolns}) e.g. 
\begin{equation}
   \min(|| \vec{x}_t - \vec{x}_{t-\Delta t} ||, ~
 || \vec{x}_t - \sigma\,\vec{x}_{t-\Delta t} ||)
   ~~~\mbox{or}~~~
   \min_l(|| \vec{x}_t - g(-l)\vec{x}_{t-\Delta t} ||)~.
\end{equation}
Minimisation over shifts can be avoided if slicing is applied
(section \ref{sect:slicing}).

The norm itself might need tinkering with.  For example, in the 
sheared flow of fluid, perturbations in the streamwise dimension
are typically an order of magnitude larger than the crossflow components.
A `compensatory' norm helps in this case, where the components are scaled to be more similar in magnitude.

\section{The Newton--Krylov method}

The Jacobian-free Newton--Krylov (JFNK) method is a variant 
of the Newton--Raphson method.  In its raw form, the Newton--Raphson
method for an $n$-dimensional system involves an $n\times n$ 
Jacobian matrix, which can be tricky to evaluate.  It is possible
to avoid this evaluation using a Krylov-subspace method
\citep{knoll2004jacobian}.

\subsection{The Newton--Raphson method}

To find roots $x$ such that $f(x)=0$ in one dimension, 
given an initial guess $x_0$, the Newton-Raphson method 
generates improvements using the iteration
\begin{equation}
   x_{i+1} = x_i - f(x_i)/f'(x_i) \, .
\end{equation}
Re-arranging,
we may re-express the iteration as 
\begin{equation} \label{eq:NR1d}
   x_{i+1} = x_i + \delta x_i \, 
   ~~~\mbox{where}~~~ 
   f'(x_i)\,\delta x_i = -f(x_i) \, .
\end{equation}
Our task is to find fixed points of the map 
such that $\vec{x_p}=\vec{\FM}(\vec{x}_p)$, i.e.\
\begin{equation}
   \vec{F}(\vec{x}_p)=\vec{0}
   ~~~\mbox{where}~~~ 
   \vec{F}(\vec{x}) = \vec{\FM}(\vec{x}) - \vec{x} .
\end{equation}
(The fixed points could correspond to equilibria, periodic
orbits, or their relative equivalents. Augmentations, if necessary,
to find a period $T$ or spatial shift $l$ are delayed to
section \ref{sect:augTl}.)
The extension of Newton's method (\ref{eq:NR1d})
to an $n$-dimensional system is then
\begin{equation} \label{eq:N-R}
   (a)~~\vec{x}_{i+1} = \vec{x}_i + \vec{\delta x}_i \, 
   ~~~\mbox{where}~~ 
   (b)~~\left. \frac{\vec{\partial F}}{\vec{\partial x}}\right|_{\vec{x}_i}
\vec{\delta x}_i = -\vec{F}(\vec{x}_i) \, .
\end{equation}
In order to apply the update (\ref{eq:N-R}a), the linear system (\ref{eq:N-R}b) needs to be solved for the unknown $\vec{\delta x}_i$.

In (\ref{eq:N-R}b), the matrix part is given by 
\begin{equation}
   \left.\frac{\vec{\partial F}}{\vec{\partial x}}\right|_{\vec{x}_i}
   ~~=~~ 
   \left.\frac{\vec{\partial \FM}}{\vec{\partial x}}\right|_{\vec{x}_i}
   - I
   ~~=~~ J - I
\end{equation}
where $J$ is the {\bf Jacobian} matrix for $\vec{\FM}(\vec{x})$
and $I$ is the identity matrix.
For the case $n=3$,
\begin{equation}
  \vec{x} = (x_1,\,x_2,\,x_3)\,,
  \quad
  \vec{\FM}(\vec{x})=\left[\begin{array}{c}
     \FM_1\\ \FM_2\\ \FM_3
     \end{array}\right],
  \quad
  J = 
  \left[\begin{array}{ccc}
     \frac{\partial \FM_1}{\partial x_1} &
     \frac{\partial \FM_1}{\partial x_2} &
     \frac{\partial \FM_1}{\partial x_3} \\[2pt]
     \frac{\partial \FM_2}{\partial x_1} &
     \frac{\partial \FM_2}{\partial x_2} &
     \frac{\partial \FM_2}{\partial x_3} \\[2pt]
     \frac{\partial \FM_3}{\partial x_1} &
     \frac{\partial \FM_3}{\partial x_2} &
     \frac{\partial \FM_3}{\partial x_3} 
  \end{array}\right] \, .
\end{equation}

\subsection{Jacobian-Free method}
\label{sect:JFmethod}

The $n\times n$ Jacobian matrix $J$ is usually difficult to evaluate. 
We might not even have sufficient computer memory to store it for a high 
dimensional system.
The problem (\ref{eq:N-R}b), however, is in the form 
\begin{equation}
 A\,\vec{\delta x}=\vec{b}\,,
\end{equation}
where $A$ is an $n\times n$ matrix and $\vec{\delta x}$ and $\vec{b}$
are $n$-vectors.  This can be solved for $\vec{\delta x}$ using the 
{\bf Krylov-subspace method $\mbox{GMRES(m)}$}.  
The GMRES algorithm does not need to know the matrix $A$ itself, only the result of multiplying a given vector by $A$.  
The method seeks a solution for $\vec{\delta x}$ in 
$\mathrm{span}\{\vec{K}_1,\,\vec{K}_2,\dots,\,\vec{K_m}\}$,
i.e.\
$
\vec{\delta x} = c_1\,\vec{K}_1 + c_2\,\vec{K}_2 + ... +c_m\,\vec{K}_m
$.
It is common to start with $\vec{K}_1=\vec{b}/||\vec{b}||$.
The next vector is generated by evaluating $\tilde{\vec{K}}_{i+1}=A\,\vec{K}_i$, then $\vec{K}_{i+1}$ is obtained by orthonormalising 
$\tilde{\vec{K}}_{i+1}$
against the previous $\vec{K}_j$ ($j\le i$)
using the Gram-Schmidt method.  Next, 
$  \mathrm{error} = || A~\vec{\delta x} - \vec{b} \, ||$
is minimised over the coefficients $c_j$ ($j\le i+1$)
and the process repeated if $\mathrm{error}$ is too large.

Iterations of the GMRES algorithm for the problem (\ref{eq:N-R}b) involve calculating matrix-vector products 
with given $\vec{\delta x}$ that may be approximated:
\begin{equation}\label{eq:Japprox}
 \left.\frac{\vec{\partial F}}{\vec{\partial x}}\right|_{\vec{x}_i}\,\vec{\delta x} ~~\approx~~
  \frac{1}{\epsilon}\,
  (\vec{F}(\vec{x}_i+\epsilon\,\vec{\delta x})-\vec{F}(\vec{x}_i)) \,.
\end{equation}
$\epsilon$ is a small scalar value; a typical value is $\epsilon$ such that $(\epsilon||\vec{\delta x}||)\,/\,||\vec{x}_i||=10^{-6}$.  The important point is that we do not need to know the Jacobian --- {\bf only a routine for evaluating $\vec{F}(\vec{x})$ is required}.

Note that provided that each step of the Newton method, $\vec{\delta x}$, 
takes $\vec{x}$ in approximately the correct direction, the method is expected to converge.  Therefore the tolerance specified in the accuracy of the solution for $\vec{\delta x}$ in each Newton step (calculated via the GMRES method) typically need not be so stringent as the tolerance placed on the Newton method itself for the solution $\vec{x}$.  For example, we might seek a relative error for the Newton solution $||\vec{F}(\vec{x})||/||\vec{x}||=O(10^{-8})$, but a relative error for the GMRES solution $||A\,\vec{\delta x}-\vec{b}||/||\vec{\delta x}||=O(10^{-3})$ is likely to be sufficient for calculation of the steps $\vec{\delta x}$.

\subsection{Hookstep approach}
\label{sect:hookstep}

\begin{figure}[]
 \centering
 \vspace{-5mm}
 \includegraphics[width=70mm]{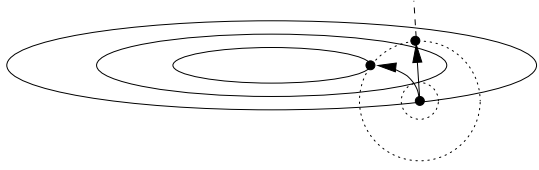}
 \caption{ \label{fig:hookstep}
 Hookstep versus `damped'/line-search step of the same size
 in minimising $||\vec{F}(\vec{x})||^2$.
 The radius of the circle corresponds to the size of the step / 
 trust region $\delta$.
 }
\end{figure}

To improve the domain of convergence of the Newton method, it is commonplace to limit the size of the step taken.  One approach is simply to take a `damped' step in the direction of the solution to \ref{eq:N-R}(b), i.e.\ step by $\alpha\ \vec{\delta x}_i$, where $\alpha \in (0,1]$.  In the '''hookstep approach''', we minimise subject to the condition that the magnitude of the Newton step is limited, $||\vec{\delta x}_i||<\delta$, where $\delta$ is the size of the '''trust region''':
\begin{equation} \label{eq:dxerr}
\min_{\vec{\delta x}_i:\ ||\vec{\delta x}_i||<\delta} \  \left|\left|\left. \frac{\vec{\partial F}}{\vec{\partial x}}\right|_{\vec{x}_i}
\vec{\delta x}_i + \vec{F}(\vec{x}_i) \right|\right|\ .
\end{equation}
Given the minimisation, the hookstep $\vec{\delta x}_i$ is expected to produce a better result than a simple damped step of the same size.  It is also expected to perform much better in 'valleys', where it produces a bent/hooked step to a point along the valley, 
rather than jumping from one side of the valley to the other;
see figure \ref{fig:hookstep}.

The hookstep can be calculated with little extra work to the GMRES method, provided that the size of Krylov-subspace, m, is chosen sufficiently large to solve to the desired accuracy within m GMRES iterations; for details
see \cite{Visw07b} [particularly {\bf v1} on arxiv.org].

For a given $\vec{\delta x}_i$, the reduction in error predicted by the linearisation (\ref{eq:dxerr}) can be compared with the actual reduction in $||\vec{F}(\vec{x}_i+\vec{\delta x}_i)||$.  According to the accuracy of the prediction, the size of the trust region $\delta$ can be adjusted automatically; see \cite{Dennis96}.

\subsection{Adding constraints}
\label{sect:augTl}

\vspace{-2mm}
\subsubsection{Time constraint}
\vspace{-2mm}

When looking for a periodic orbit, the period $T$ is an 
extra unknown.  One way to eliminate needing to find $T$ is 
to work within a Poincar\'e section, as described in section
\ref{sect:Poincare}.  We can then attempt to solve the function 
$\vec{F}(\vec{x})=\vec{\FM}(\vec{x}) - \vec{x}$ 
as it stands to find a point $\vec{x}_p$ on the Poincar\'e 
section that corresponds to a periodic orbit.

We might not want to restrict ourselves to Poincar\'e sections.
We must then solve
\begin{equation}
  \vec{F}(\vec{x},T) = \vec{\FM}^T(\vec{x})-\vec{x} = \vec{0} \, ,
\end{equation}
for $(\vec{x},T)$.  We augment the whole system.  Let
\begin{equation} \label{eq:augTb}
   \tilde{\vec{x}}_i = (\vec{x}_i,\,T_i) \,
   ~~~\mbox{and}~~~ \tilde{\vec{b}} = (-\vec{F}(\tilde{\vec{x}}_i),\,0) \, .
\end{equation}
We now want to solve a system of the form
\begin{equation}
   \label{eq:augT}
   A\,\tilde{\vec{\delta x}}_i = \tilde{\vec{b}}\, ,
\end{equation}
for $\tilde{\vec{\delta x}}_i= (\vec{\delta x}_i,\,\delta T_i)$, but need an extra constraint because we have an extra unknown.  We choose that the update 
$\vec{\delta x}_i$ has no component that points along the
trajectory, 
i.e.~$\langle\dot{\vec{x}_i}|\vec{\delta x}_i\rangle=0$.
Following the ethos of matrix-free methods, that we do not need to know the matrix $A$ itself, we only need to state the result of 
multiplication by $A$:
\begin{equation} \label{eq:multJaugT}
   A\,\tilde{\vec{\delta x}} = 
   \left(
    \left.\frac{\vec{\partial F}}{\vec{\partial \tilde{x}}}\right|_{\tilde{\vec{x}}_i}\,\tilde{\vec{\delta x}}
   ,\, \langle\dot{\vec{x}_i}|\vec{\delta x}\rangle
   \right) \,  .
\end{equation}
We use the approximation (\ref{eq:Japprox}) 
to evaluate the first part
of the result.
The augmented system (\ref{eq:augT}) can be solved 
for $\tilde{\vec{\delta x}}_i$ using the GMRES algorithm
by applying multiplications (\ref{eq:multJaugT}). 
The update for both the state and the period is then
$\tilde{\vec{x}}_{i+1}=\tilde{\vec{x}}_i+\tilde{\vec{\delta x}}_i$.

\subsubsection{Shift constraints}

For relative equilibria (travelling waves) 
and relative periodic periodic orbits we need to solve
\begin{equation}
   \vec{F}(\vec{x},T,l) = g(-l)\,\vec{\FM}^T(\vec{x})-\vec{x} = \vec{0}\, ,
\end{equation}
where $l$ is an unknown spatial shift in the homogeneous $x$-dimension.

One way to avoid the extra unknown is to work with the sliced
dynamics (section \ref{sect:slicing}) so that 
arbitrary shifts are automatically eliminated.
Alternatively, we can augment the system again.  Let
\begin{equation}
   \tilde{\vec{x}}_i = (\vec{x}_i,\,T_i,l_i) \,
   ~~~\mbox{and}~~~ \tilde{\vec{b}} = (-\vec{F}(\tilde{\vec{x}}_i),\,0,\,0) \, .
\end{equation}
We now want to solve a system of the form
\begin{equation}
   \label{eq:augTl}
   A\,\tilde{\vec{\delta x}}_i = \tilde{\vec{b}}\, ,
\end{equation}
for $\tilde{\vec{\delta x}}_i$, but need another constraint 
to match the extra unknown.  This time we choose that 
the update $\vec{\delta x}_i$ has no component that just 
corresponds to a spatial shift, 
i.e.~$\langle\partial_x\vec{x}_i|\vec{\delta x}_i\rangle=0$.
We assert that multiplication by $A$ is:
\begin{equation}
   A\,\tilde{\vec{\delta x}} = 
   \left(
    \left.\frac{\vec{\partial F}}{\vec{\partial \tilde{x}}}\right|_{\tilde{\vec{x}}_i}\,\tilde{\vec{\delta x}}
   , \, 
   \langle\dot{\vec{x}_i}|\vec{\delta x}\rangle
   , \,
   \langle\partial_x\vec{x}_i|\vec{\delta x}\rangle
   \right) \,  .
\end{equation}

\subsection{Preconditioning}
\vspace{-2mm}

The good news is that you can ignore preconditioning 
and skip this section if you are combining 
Newton--Krylov with timestepping to evaluate the flow-map
$\vec{\FM}^T(\vec{x})$.

\vspace{-2mm}
\subsubsection{Exponentiation and timestepping}
\vspace{-2mm}

The GMRES algorithm is closely related to another
Krylov-subspace method, the Arnoldi method,
which is used to calculated eigenvalues of a matrix $A$.
It tends to find the eigenvalues most separated in the complex plane first,
but those might be of little interest.  For example, the Laplacian
$\nabla^2$ has a spectrum of very negative eigenvalues corresponding
to high frequency oscillations that rapidly decay.  Basically,
we do not
wish to build a Kyrlov-subspace involving such modes.

It may be better to work with 
$\tilde{A}=\mathrm{e}^A=1+A+\frac{1}{2!}A^2+...$, corresponding
to the eigenproblem 
$\mathrm{e}^\sigma\vec{x}=\mathrm{e}^A\vec{x}$.  This problem
shares the same eigenvectors as the problem $\sigma\,\vec{x}=A\,\vec{x}$, but has more suitable eigenvalues, $\tilde{\sigma}=\mathrm{e}^\sigma$. The negative eigenvalues 
$\sigma$ then correspond to eigenvalues $\tilde{\sigma}$ bunched close to the origin.  The Arnoldi method then 
favours the $\tilde{\sigma}$ most distant from the origin, corresponding
to the $\sigma$ with largest real parts.

Note that for the system $\partial_t \vec{x} = A\,\vec{x}$, time integration corresponds to exponentiation: Taking eigenvector $\vec{x}$ with growth rate $\sigma$ as an initial condition, the result of time integration from $0$ to $T$ is $\mathrm{e}^{\sigma T}\vec{x}$.  We therefore have that $\mathrm{e}^{\sigma T}\vec{x}=\vec{x}+\int_0^T A\,\vec{x}\,dt=\mathrm{e}^{AT}\vec{x}$, 
which can be written 
$\tilde{\sigma}\,\vec{x}=B\,\vec{x}$ where
$\tilde{\sigma}=\mathrm{e}^{\sigma T}$ is the eigenvalue of the time integration operator $B=\mathrm{e}^{AT}$.


\subsubsection{Explicit preconditioning}

GMRES is likely to find it easier to solve $M^{-1}A\,\vec{x}=M^{-1}\vec{b}$ than the original system, if $M^{-1}$ is an approximate inverse for $A$.
For example, if $A$ is dominated by its diagonal elements, we might take $M$ to be the banded matrix consisting of the diagonal and the first sub- and super-diagonals of $A$.  Each GMRES iteration applied to the modified system now requires a muliplication by $A$ then by $M^{-1}$.  This is fine, as, 
for a banded matrix, it is quick and easy to solve $M\vec{x}'=\vec{x}$ for $\vec{x}'$.  
Like $A$, we don't need to know the matrix $M^{-1}$ itself, only the result of multiplication by the matrix.

\vspace{5mm}
\section{Try it yourself! 
 Application of the Newton--Krylov method 
 to the Lorenz system}

Given that the Newton--Krylov method is designed to cope with 
high-dimensional systems (the same code has been used to find
travelling waves in pipe flow), this is somewhat overkill, 
but it helps illustrates how we can use the solver as a 
{\bf black box}...
\\

{\bf Please cite \href{http://www.openpipeflow.org}{openpipeflow.org} \citep{Willis17} if you use this
code in your research. Thanks!}\\

\begin{itemize}
\setlength\itemsep{10pt}
\item
{\bf Download the Template/Example}
(Fortran90 / MATLAB\,/\,Octave) \\
\url{http://www.openpipeflow.org/index.php?title=Newton-Krylov_method}
\item
For MATLAB, the unpacked \texttt{tgz}/\texttt{zip} file has separate \texttt{.m} 
files for each function.  \\
{\bf Take a look at}
  \begin{itemize} 
  \item \texttt{Lorenz\_f.m}: Lorenz evolution rule 
     $\dot{\vec{x}}=\vec{f}(\vec{x})$.
  \item \texttt{steporbit.m}: Evaluate $\vec{\FM}^T(\vec{x})$, i.e. step $\vec{f}$ by \texttt{ndts\_}
     timesteps, where the input \texttt{x(1)}$=T$, 
     and \texttt{x(2:4)}$=(X,Y,Z)$.  The timestep size is 
     \texttt{dt}$=T/$\texttt{ndts\_}.
  \item \texttt{saveorbit.m}: Output at end of each Newton iteration.
 \texttt{relative\_err}$\,=||\vec{F}(\vec{x})||\,/\,||\vec{x}||$.
  \item \texttt{MAIN.m}:  Set up initial guess $\vec{x}_0$ and 
     call the {\bf black box} \texttt{NewtonHook.m}.
  \end{itemize}
\newpage
\item
Other functions are called by 
\texttt{NewtonHook.m}, and are unlikely
to need changing for a problem of this type,
where shifts and other spatial symmetries are ignored:
  \begin{itemize} 
  \item \texttt{getrhs.m}: Evaluate right-hand side $\tilde{\vec{b}}$ 
  (\ref{eq:augTb}) i.e.~$\vec{F}(\tilde{\vec{x}})=\vec{\FM}^T(\vec{x})-\vec{x}$.
  \item \texttt{multJ.m~}: Evaluate multiplication (\ref{eq:multJaugT}),
  i.e.~multiplication by the Jacobian.
  \item \texttt{multJp.m}: Preconditioner for multiplication 
  (here an empty function).
  \item \texttt{dotprd.m}: Evaluate inner product 
      $\langle\vec{a}|\vec{b}\rangle$.
  \item \texttt{GMRESm.m}: Method of section \ref{sect:JFmethod}.
  \item \texttt{GMREShook.m}: Calculate hookstep, section \ref{sect:hookstep}.
  \end{itemize}

\item
The following data are points on the periodic orbits of figure 
\ref{fig:vis03}, taken from \cite{DV03}.  $Z=27$ in call cases.
\begin{center}
{\small \begin{tabular}{llll}
\hline
 & $X$ & $Y$ & $T$ \\
\hline
AB & −13.763610682134 & −19.578751942452 & 1.5586522107162 \\
AAB & −12.595115397689 & −16.970525307084 & 2.3059072639399 \\
AAAB & −11.998523280062 & −15.684254096883 & 3.0235837034339 \\
AABB & −12.915137970311 & −17.673100172646 & 3.0842767758221 \\
\hline & & ($Z=27$)
\end{tabular} }
\end{center}
\item
In MATLAB, call \texttt{MAIN}.
It will plot the result of timestepping the initial guess for the 
AB orbit (green), call the \texttt{NewtonHook} subroutine, then
plot the converged solution (blue).
Scroll back through the output, and 
compare \texttt{relative\_err} 
for the initial guess at \texttt{iteration 0}
with the final relative error. 
\item
Comment/uncomment other initial guesses \texttt{new\_x}$\,=\vec{x}_0$,
or experiment with your own.  How do they affect the number of 
Newton iterations taken? \newline
[Typically convergence takes $O(10)$ iterations, otherwise it will never converge.]
\item
Uncomment the initial
guess for an equilibrium.  Here we assume a short fixed $T$, too 
short for a PO; $T$ is not permitted to change, otherwise 
$||\vec{\FM}^T(\vec{x})-\vec{x}||$ could be reduced by simply taking $T\to0$.
Check that \texttt{MAIN} can find the 
analytic equilibrium solution $(\pm\alpha,\pm\alpha,r-1)$, 
where $\alpha=\sqrt{(r-1)\,b}$. 
\end{itemize}

\subsection{Adapting the code for your own use}

\begin{itemize}
\setlength\itemsep{10pt}
\item
For a very large system, 
for which you might consider parallelization 
(see final comment),
you should probably use the Fortran90 version.
\item
Experiment with the Template/Example first, to get used to 
how the code is set up.  The initial guess is put in \texttt{new\_x}.
\item
Note that at present, \texttt{new\_x(1)}$\,=T$ (the period), 
and \texttt{new\_x(2:end})$\,=\vec{x}$ (the state).
\item
The place to start is then \texttt{steporbit}.  If you already have 
an existing timestepping code, it could do something as simple 
as call it externally via system calls:
\newpage
{\small
\begin{verbatim}
 function y = steporbit(ndts_,x)
   persistent dt

   if ndts_ ~= 1                % Set timestep size dt=T/ndts_
      dt = x(1) / ndts_ ;       % If only doing one step to calc \dot{x},
   end                          % then use previously set dt.

   a = x(2:end) ;

   WRITE DATA TO FILES:
      dt     timestep size
      ndts_  number of steps to take
      a      initial condition
   
   LOAD STATE, TIMESTEP, SAVE STATE:
      system('run_my_code.exe')   
   
   LOAD TIMESTEPPED STATE: --> a

   y = zeros(size(x)) ;
   y(2:end) = a ;
 end 
\end{verbatim}
}

\item
\texttt{saveorbit} is called at the end of each Newton iteration.
Add code here to save the current state \texttt{new\_x}.

\item
If your inner product corresponds to 
$\langle\vec{a}|\vec{b}\rangle \,=\, \vec{a}^TW\vec{b}$ 
where $W$ is a diagonal matrix of positive weights, and here $T$ is the 
transpose, then pass $\vec{x}'=W^\frac{1}{2}\vec{x}$ to the code.
The existing functions that take inner products then need no modification.

\item
For parallel use with MPI+Fortran, the \texttt{NewtonHook} and
\texttt{GMRES} codes do not need changing:
Split vectors over threads and let each thread pass its section
to \texttt{NewtonHook}.
The only place where an MPI call is required is an MPI\_Allreduce in the 
\texttt{dotprod} function.
To avoid all threads outputting information, set \texttt{info=1}
on rank 0, and \texttt{info=0} on all other ranks.

\item
Further information at \href{http://www.openpipeflow.org}{openpipeflow.org}.
\end{itemize}

\vspace{10mm}
{\Large \bf Acknowledgements}
\vspace{2mm}

AW would like to thank 
Rich Kerswell, 
Predrag Cvitanovi\'c (\href{http://www.chaosbook.org}{chaosbook.org}), 
John Gibson (\href{http://www.channelflow.org}{channelflow.org}), 
Marc Avila
and many others for their generous support in many forms.
Developed under EPSRC grants EP/K03636X/1, EP/P000959/1.

\newpage
\bibliographystyle{chicago}
\bibliography{pipes.bib}
\end{document}